\documentclass[preprint]{vgtc}               

\ifpdf
  \pdfoutput=1\relax                   
  \usepackage{graphicx}                
  \DeclareGraphicsExtensions{.pdf,.png,.jpg,.jpeg} 
\else
  \usepackage{graphicx}                
  \DeclareGraphicsExtensions{.eps}     
\fi%

\graphicspath{{figures/}{pictures/}{images/}{./}} 

\usepackage{microtype}                 
\PassOptionsToPackage{warn}{textcomp}  
\usepackage{textcomp}                  
\usepackage{mathptmx}                  
\usepackage{times}                     
\usepackage{cite}                      
\usepackage{tabu}                      
\usepackage{booktabs}                  
\usepackage{enumitem}
\usepackage[skip=1pt]{caption}
\usepackage{amsmath}
\usepackage{balance}
\newcommand{\system}{\textit{GreenMine}}

\usepackage{xcolor}
\usepackage{soul}
\usepackage{array}
\usepackage{tabularray}
 \definecolor{sampink}{HTML}{ff0000}
 \definecolor{sampink}{HTML}{000000}
 \newcommand{\sam}[1]{{\color{sampink} #1}}
 
\usepackage{marginnote} 

\renewcommand*{\marginnote}[1]{}
\makeatletter
\def\input@path{{sections/}}

\onlineid{7280}

\vgtccategory{Research}

\vgtcinsertpkg

\ieeedoi{https://doi.org/10.1109/PacificVis64226.2025.00037}

\title{Visual Text Mining with Progressive Taxonomy Construction\\ for Environmental Studies}

\author{Sam Yu-Te Lee\thanks{e-mail: ytlee@ucdavis.edu} %
\affiliation{\scriptsize University of California, Davis}
\and Cheng-Wei Hung\thanks{e-mail: d10541002@ntu.edu.tw}
\affiliation{\scriptsize National Taiwan University}
\and Mei-Hua Yuan\thanks{e-mail: meihuayuan@gate.sinica.edu.tw}
\affiliation{\scriptsize Academia Sinica}
\and Kwan-Liu Ma \thanks{e-mail: klma@ucdavis.edu} %
\affiliation{\scriptsize University of California, Davis}
}

\abstract{
Environmental experts have developed the DPSIR (Driver, Pressure, State, Impact, Response) framework to systematically study and communicate key relationships between society and the environment. Using this framework requires experts to construct a DPSIR taxonomy from a corpus, annotate the documents, and identify DPSIR variables and relationships, which is laborious and inflexible. Automating it with conventional text mining faces technical challenges, primarily because the taxonomy often begins with abstract definitions, which experts progressively refine and contextualize as they annotate the corpus. In response, we develop \system, a system that supports interactive text mining with prompt engineering. The system implements a prompting pipeline consisting of three simple and evaluable subtasks. In each subtask, the DPSIR taxonomy can be defined in natural language and iteratively refined as experts analyze the corpus. To support users evaluate the taxonomy, we introduce an uncertainty score based on response consistency. Then, we design a radial uncertainty chart that visualizes uncertainties and corpus topics, which supports interleaved evaluation and exploration. Using the system, experts can progressively construct the DPSIR taxonomy and annotate the corpus with LLMs. Using real-world interview transcripts, we present a case study to demonstrate the capability of the system in supporting interactive mining of DPSIR relationships,
and an expert review in the form of collaborative discussion to understand the potential and limitations of the system. We discuss the lessons learned from developing the system and future opportunities for supporting interactive text mining in knowledge-intensive tasks for other application scenarios.
} 

\CCScatlist{
  \CCScatTwelve{Information systems}{In\-for\-mation sys\-tems app\-li\-ca\-tions}{Data Min\-ing}{};
  \CCScatTwelve{Human-centered computing}{Visu\-al\-iza\-tion}{Visu\-al\-iza\-tion sys\-tems and tools}{}
}

\begin{document}
\maketitle

\section{Introduction}
Human activities have a profound influence on the environment, which, in turn, shape policy-making and human behavior.
The DPSIR framework (\textbf{D}river, \textbf{P}ressure, \textbf{S}tate, \textbf{I}mpact, \textbf{R}esponse) is commonly used in environmental science to study and communicate the complex relationships between societal and environmental factors. 
Environmental experts would collect documents and manually mine for insights on DPSIR. 
As the corpus scales, this process quickly becomes laborious and inefficient. 

Such a task can be formulated as a text mining operation and automated. 
The definitions of DPSIR essentially form a ``label taxonomy''.
Conventionally, the taxonomy can be constructed manually or semi-automatically with clustering techniques~\cite{wan2024textminingllm} and then used to annotate the corpus with pre-trained models.
\sam{
\marginnote{$\triangle$\_1\_1}However, mining with the DPSIR framework is a progressive process. 
The DPSIR taxonomy often starts with abstract definitions and is gradually enriched as experts investigate the corpus. The investigation usually contains two iterative steps: first, each DPSIR indicator is contextualized with ``variables'', such as population or pollution, which constitute the taxonomy; second, relationships between variables are constructed with supporting evidence (e.g., text snippets) from the corpus and documented as insights. 
Then, experts would investigate the corpus again with new insights in mind and repeat this process until no new insight is derived.
Due to this progressive characteristic, it is impractical to construct validation sets with human annotations to validate the taxonomy and insights in each iteration.
This unique challenge motivates us to utilize Large Language Models (LLMs) to support experts in this progressive process and use uncertainty metrics to support expert validation. 
With LLMs, the DPSIR taxonomy can be expressed in natural language and then inserted in expert prompt templates~\cite{macneil2023promptmiddleware}, which instruct LLMs to mine mentions of DPSIR variables and relationships from the corpus.
This allows experts to progressively refine and contextualize the DPSIR taxonomy as they explore the corpus and gain new insights.
}

Building upon this approach, we work closely with environmental experts and identify four technical challenges.
First, designing prompts is not trivial for inexperienced users~\cite{zamfirescu2023johnny, kim2024evallm}, especially when the prompt is applied to a dataset~\cite{lee2024awesum}.
Second, mining DPSIR insight from a corpus in a single zero-shot prompt is still too complex even for state-of-the-art AI models (e.g., GPT-4). The mining needs to be decomposed into a prompting pipeline to get satisfactory results.
Third, evaluation is still necessary to ensure the reliable performance of the prompt, which is especially challenging without ground truths.
Fourth, the mined insights are scattered across the corpus in textual form, which is not ideal for communication. 

In this work, we address these challenges and develop \system, an LLM-based system that supports human-in-the-loop text mining with the DPSIR framework and visualization of the results.
To lower the technical barrier, the system takes expert input for the DPSIR taxonomy in natural language and inserts them into prompt templates for execution.
For more accurate results, the system decomposes the mining task into three sequential subtasks that are more manageable and sensible.
\sam{
We design an uncertainty chart that visualizes the uncertainty of the LLM responses and the topic distribution in the supporting evidence.
}
Finally,  to support communication, we design a DPSIR graph that organizes the mining results in a graph that can be interactively explored.
We present a case study to show the effectiveness of the LLM-based mining support, 
and an expert review to discuss the system's potential and limitations.
We found that while the system is designed for environmental studies, the technical challenges and solutions have broader applications. We discuss lessons learned and future opportunities for supporting human-in-the-loop text mining in other knowledge-intensive scenarios.

We consider our contributions are:
\begin{itemize}[noitemsep, topsep=0pt]
    \item introducing a prompt-based text mining pipeline for human-in-the-loop DPSIR mining from a corpus,
    \item designing a visual analytics system integrating the mining pipeline with uncertainty charts for prompt refinement and a DPSIR graph for collaborative discussion, and
    \item providing lessons learned from the design process that inform other applications with knowledge-intensive tasks.
\end{itemize}

\vspace*{-0.15cm}
\section{Background}
\sam{
\marginnote{$\triangle$\_2\_1}The DPSIR framework~\cite{atkins2011dpsir} provides a holistic approach to studying the connections that exist within and between diverse and complex societal and environmental factors, which is critical to support policymakers in their decision-making.}
According to the framework, \textit{Drivers} are the key demands by society and create \textit{Pressures} on the environment.
 Accumulation of pressures leads to \textit{State} changes, which creates \textit{Impacts} on human society or the environment. Significant impacts may elicit a societal \textit{Response} and alter the \textit{Drivers} and \textit{Pressures}, forming a feedback loop. 
We refer to them as \textit{indicators}. 

Given the diversity and complexity of the indicators, applying the DPSIR framework to study a specific region requires standardization and contextualization of the definitions of each indicator into \textit{variables}~\cite{oesterwind2016dpsiruntangle}, so that their relationships can be studied. 
For example, to study the reef fishing activities in Kenya, Mangi et al.~\cite{mangi2007reefdpsir} delineates \textit{Drivers} into several variables such as population, unemployment, tradition, and culture. Such a delineation is highly dependent on the regional context, making it infeasible to propose a taxonomy that is generally applicable. Environment experts need to manually build the taxonomy from large collections of documents, such as prior research papers, reports, or interview transcripts.
From a text mining perspective, the variables are concepts to be extracted, and the mining goal is to progressively add concepts to extract and uncover relationships between concepts.

In this work, our collaborating environmental experts apply the DPSIR framework to a study conducted in Lyudao (also known as Green Island), Taiwan, a small island in the Pacific Ocean. They have derived an initial taxonomy from a literature review, and have conducted interviews with the local residents. Using the interview transcripts as a corpus, they seek to contextualize and refine the taxonomy, and then mine key relationships between the variables. 

\begin{figure}[t]
  \centering
  \includegraphics[keepaspectratio, width=0.95\columnwidth]{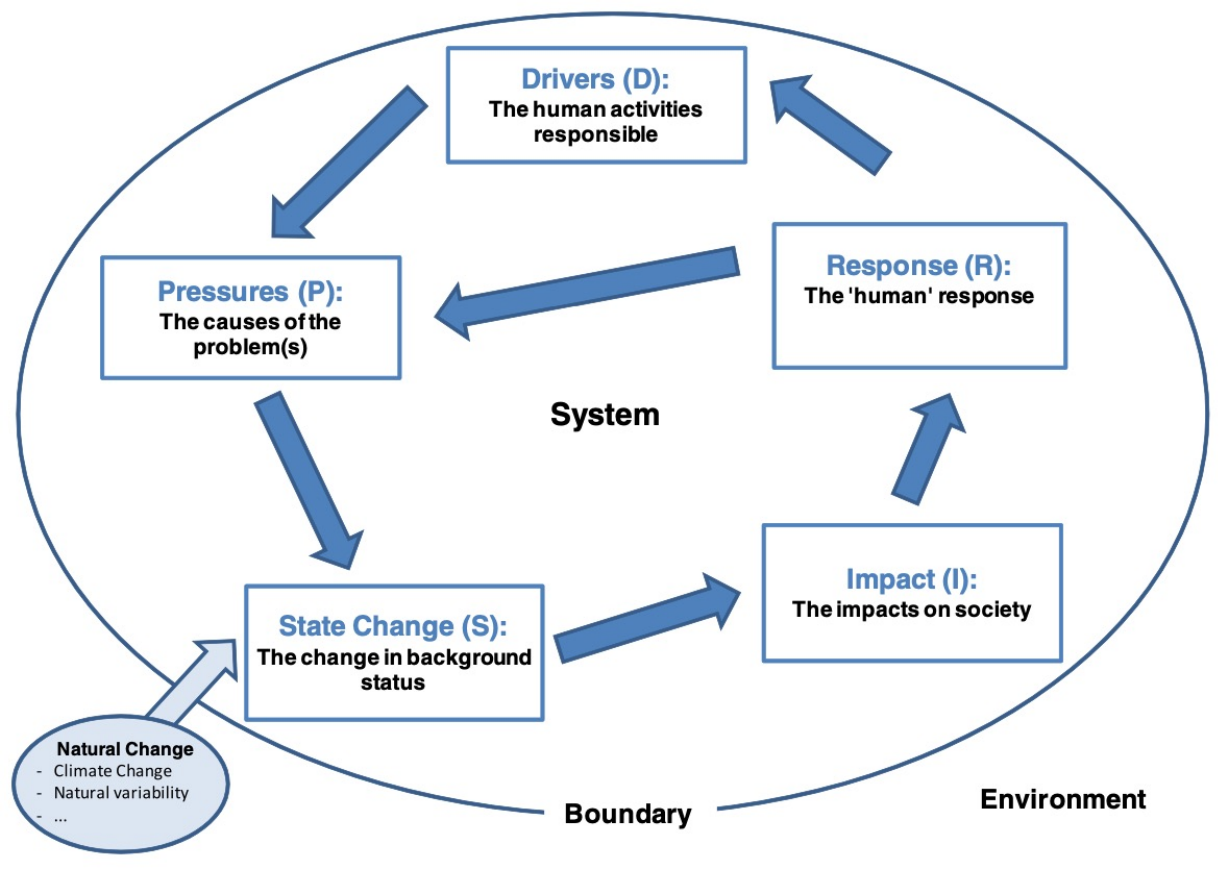}
  \caption{Example of a DPSIR diagram from Atkins et al.~\cite{atkins2011dpsir}. The DPSIR framework is commonly depicted as a cyclic feedback loop between indicators. Our radial DPSIR graph design extends the cyclic characteristic with progressive disclosure.}
  \label{fig: dpsir_diagram}
  \vspace*{-0.5cm}
\end{figure}

The DPSIR framework also facilitates communication of the findings to policymakers. Similar to business strategy diagrams~\cite{brath2024managementdiagram}, there are established infographic diagrams for the DPSIR framework familiar to the experts and the policymakers. The diagram is often organized in a loop that connects each indicator to reflect the cyclic feedback between society and the environment, as shown in~\autoref{fig: dpsir_diagram}. In this work, we extend this infographic design with progressive disclosure to better support collaborative discussion. 

\vspace*{-0.2cm}
\section{Related Works}
Our work enhances existing visual interfaces for text mining and incorporates prompt engineering to support mining with the DPSIR framework.
We review interfaces for conventional and prompt-based text mining, and design studies for interactive prompt engineering.

\vspace*{-0.15cm}
\subsection{Visual Interfaces for Concept and Relation Mining}
Supporting and visualizing text mining have been extensively studied, incorporating various text mining and visualization techniques~\cite{liu2019bridging}. 
In the context of mining DPSIR indicators, approaches that extract and visualize concepts and relations are most relevant.
Earlier works use statistical models on word frequencies to mine valuable information.
For example, VOSViewer~\cite{wong2018vosviewer} supports the automatic extraction of important terms and co-occurrence relations from a corpus using statistical models. 
Later, techniques related to disambiguated entities or concepts become predominant.
FacetAtlas~\cite{cao2010facetatlas} extracts entities and relations using named entity recognition models~\cite{finkel2005ner}, with a focus on internal and external relations between entity classes. 
ConceptEVA~\cite{zhang2023concepteva} extracts concepts and their co-occurrence relations using domain-specific knowledge graphs~\cite{mendes2011dbpediaspotlight} to support customized generation of document summaries.
Both organize the mining result as a node-link diagram.
ConceptScope~\cite{zhang2021conceptscope} uses domain ontologies to extract concepts and their relations for analyzing documents and visualizes the result in a bubble treemap.

A common limitation of these approaches is that the label taxonomy is static, e.g., the domain ontologies or knowledge graphs.
An exception is ConceptVector~\cite{park2018conceptvector}, which supports user-controlled concept building with clusters of keywords before using them to analyze documents.
However, in the mining of DPSIR indicators, such techniques provide limited support for contextualizing the indicators. 
For example, depending on the research and dataset context, the keyword ``pollution'' can be a ``Driver'' that influences human activities, or a ``Pressure'' that influences environmental phenomena. 
In our work, we combine text mining with topic exploration to support the sensemaking of corpus and contextualization of DPSIR taxonomy.

\vspace*{-0.15cm}
\subsection{Prompt-based Text Mining and Evaluation}
Prompt Engineering is an emerging field that studies effective methods to control LLMs with natural language (i.e., prompts)~\cite{liu2023promptsurvey}, and has achieved state-of-the-art performance on text mining and information extraction tasks~\cite{dagdelen2024structured, xu2024llmforie}.
Moreover, studies have shown that integrating domain knowledge in prompts significantly improves the performance over general prompting methodologies~\cite{liu2025chemitryprompt}. 

Still, inexperienced practitioners can be over-reliant on LLMs, unaware of their limitations and risks~\cite{weidinger2022llmrisk}, such as producing misinformation. To mitigate this issue, NLP researchers have proposed techniques to evaluate LLM responses~\cite{chang2024surveyllm} from various perspectives, such as understanding, reasoning, or factuality. The most relevant to our work is on uncertainty (or confidence) estimation~\cite{chen2024quantifyinguncertaitny, Xiao2019quantify, kuhn2023semanticuncertainty}.
Xiao et al.~\cite{Xiao2019quantify} propose model and data uncertainties based on total variance. Kuhn et.al~\cite{kuhn2023semanticuncertainty} propose semantic uncertainty, which incorporates linguistic variances with off-the-shelf language models. More recently, Chen et al.~\cite{chen2024quantifyinguncertaitny} estimate model uncertainty by sampling multiple responses and their consistency.
In our work, we follow the consistency-based uncertainty estimation approaches and use Jaccard similarity to measure uncertainty. 
Then, we develop an uncertainty chart to combine uncertainty evaluation with topic exploration for progressive taxonomy construction.

\subsection{Design Studies for Prompt Engineering}
Despite the success in NLP, design studies for interactive prompt engineering have shown that writing prompts can be challenging for people without prompting experience or technical background~\cite{zamfirescu2023johnny, kim2024evallm}.
Zamifirescu et al.~\cite{zamfirescu2023johnny} found that non-technical people often overestimate the capability of LLMs because prompting imitates conversation with a human.
Moreover, the fast and constant progress in prompting methodologies~\cite{promptengineeringuide} makes it hard for them to utilize state-of-the-art prompting strategies.
Since some fundamental techniques such as chain-of-thought (CoT)~\cite{wei2023chainofthought} or retrieval-augmented-generation (RAG)~\cite{lewis2021rag}, researchers have proposed more advanced techniques like tree-of-thought (ToT)~\cite{yao2023treeofthought} and Hypothetical Document Embeddings (HyDe)~\cite{gao2022hyde}. 
In this work,  we address this challenge by exposing only the components that require domain knowledge integration to the users and integrating it into a prompting template under the hood.

Another challenge in prompt evaluation that Kim et al.~\cite{kim2024evallm} found in their user study is that evaluation is dynamic, i.e., people add additional evaluation criteria as they examine the outputs, making it hard to get actionable insights. Such a dynamic evaluation is inherent in DPSIR taxonomy mining, in that the environmental experts need to dynamically update the taxonomy as they gain a progressively deeper understanding of the corpus.
Our mining pipeline and uncertainty chart are designed to support such a dynamic evaluation.

\section{Design Analysis}
Combining the literature review, we work closely with two environmental experts, \textbf{E1} and \textbf{E2}, who are also co-authors and are currently leading an environmental study in Lyudao, to understand the process of mining with DPSIR taxonomy. 
\sam{
\marginnote{$\triangle$\_4\_1}While both experts seek to leverage LLMs for text mining, they have also encountered limitations in tools like ChatGPT~\cite{chatgpt} for controlling LLMs. 
}
Together, we identify four technical challenges in prompt-based text mining:

\paragraph{\textit{Designing dataset-scale prompts}} Designing prompts for mining a dataset requires a unique set of prompting techniques, as opposed to interacting with conversational agents such as ChatGPT~\cite{chatgpt} or Claude~\cite{claude}. For example, our experts have tried to insert a document in ChatGPT and ask for insights on DPSIR, but the results were not satisfying. Even on a single document, they need to engage in multiple rounds of conversation to calibrate the model's understanding of the DPSIR definitions. Furthermore, the various topics discussed in different documents would significantly alter the model's understanding of the DPSIR definitions, and the calibration needs to be repeated on every document. 
It is thus essential to provide support for dataset-scale prompting.

\paragraph{\textit{Decomposing the mining task}} 
Since the DPSIR mining task may be too complex for current LLMs to solve in one shot, a decomposition step~\cite{khot2023decomposed} is needed to break it into subtasks that better align with the model's capabilities. The resulting pipeline needs to support the iterative refinement of the DPSIR taxonomy for the environmental experts. This means that the subtasks should not only remain meaningful to the experts, enabling them to interpret the results without knowledge of complex prompting or ML techniques, but also be evaluable from a technical perspective.

\paragraph{\textit{Supporting evaluation and exploration}}
To support the progressive update of DPSIR taxonomy in the prompting pipeline, evaluation of the mining results and exploration of the corpus topics are the two key tasks to support. 
While topic exploration is a relatively well-studied task, evaluation without ground-truth labels remains a technical challenge. 
Moreover, the evaluation and topic should be presented in a way that caters to their background, as previous studies have shown how non-AI experts can misinterpret evaluation metrics and topic modeling results~\cite{liao2020questionAI, lee2017humantouch}.

\paragraph{\textit{Sharing the insights}} Communication of the DPSIR insights is an important purpose of applying the framework, but the mined insights are usually scattered across the corpus, and experts need to manually organize them into a form suitable for presentation and discussion. During the presentation, other experts might raise questions or propose potential ideas. However, the manual organization means the insights are static and not responsive to the discussion topics. The mining results should be automatically organized and presented in an interactive visualization.

In response to these challenges, we derive the following set of design requirements:
\begin{itemize}[label={}, leftmargin=9px, itemsep=1px, topsep=2px]
    \item \textbf{R1: Infuse domain knowledge input with established prompting methodologies.} 
    Our target users, the environmental experts, are not expected to be experienced prompting practitioners. The system should only expose components that require domain knowledge input for interaction, and seamlessly integrate it with established prompting methodologies. 
    \item \textbf{R2: Support exploration and evaluation simultaneously.} 
    To define and contextualize the DPSIR taxonomy, the expert needs to thoroughly understand the corpus, which can be cognitively demanding. The system should support the experts to use the current DPSIR taxonomy to reveal uncovered content, and iteratively refine the taxonomy. 
    At the same time, given a DPSIR taxonomy and the corresponding prompts, the experts should be able to evaluate the prompts to make sure the prompting results align with their intentions. 
    The exploration of the corpus topics and the evaluation of the prompts are intertwined, and the system should support these two tasks simultaneously.
    \item \textbf{R3: Prioritize recall over precision.}
     Considering the complexity of any environmental context, the validity of any insight can only be validated by human experts. The system should be designed to support a human-in-the-loop insight discovery, rather than aiming at extracting precise insights. Therefore, the system should prioritize recall over precision to ensure that the experts fully examine the corpus.
    \item \textbf{R4: Support collaborative discussion on mining results.} 
    After the analysis, the experts typically host internal discussions or workshops to ensure the rigorousness of the findings before communicating them to policymakers. An interactive visualization can offer graphics support for the discussion as well as querying functionalities to quickly validate ideas. Also, clarity and simplicity should be prioritized over information density to avoid distracting the discussion.
\end{itemize}

\noindent Next, we introduce how this set of design requirements informs our technical and design choices.

\begin{figure*}[t]
    \centering
    \includegraphics[width=\textwidth]{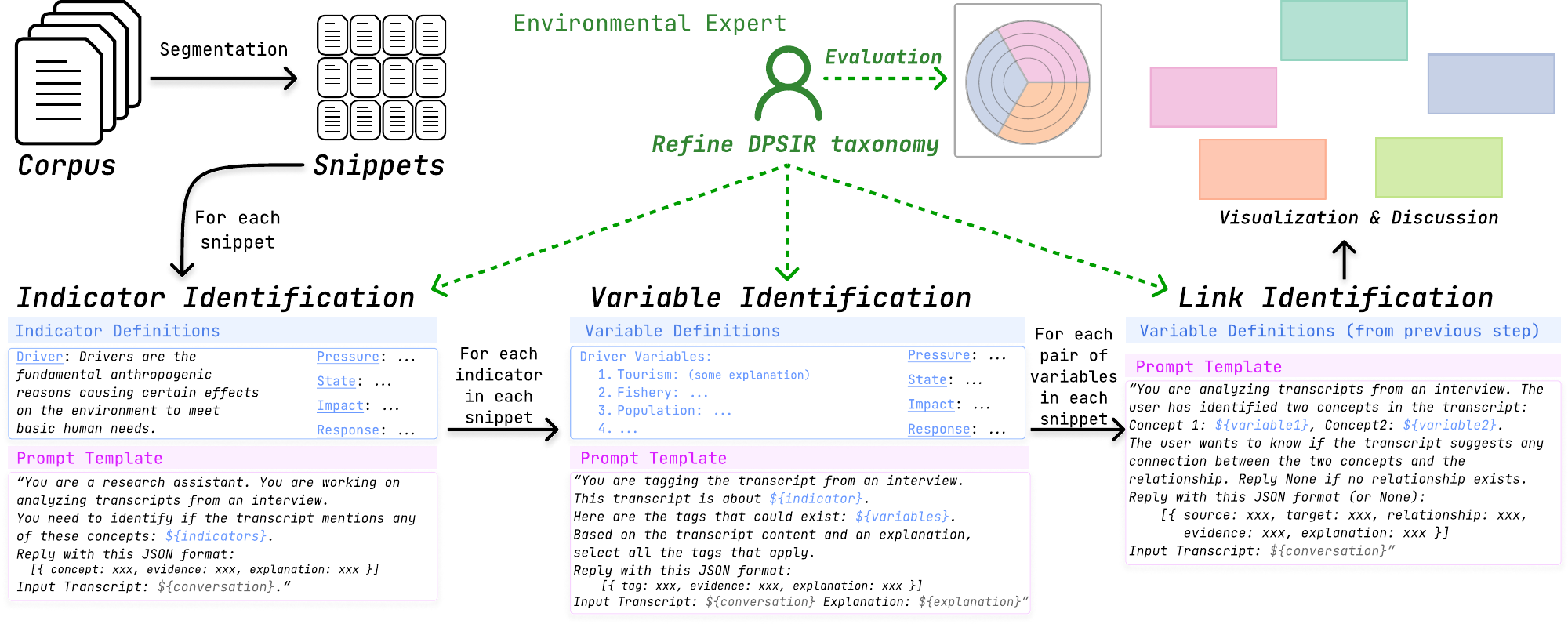}
    \caption{The system design of \system. Starting with a corpus, the documents are first segmented into snippets. Then, the snippets go through an interactive prompting pipeline consisting of three steps: Indicator, Variable, and Link Identification. In each step, the environmental experts can use the uncertainty chart to explore the corpus, evaluate the prompting performances, and refine the DPSIR taxonomy. After the mining, an interactive DPSIR Graph can be used for collaborative discussion and communication of the findings.}
    \label{fig: pipeline}
  \vspace*{-0.3cm}
\end{figure*}

\section{\system: System Design}
Informed by the design requirements, we develop \system\footnote{\url{https://github.com/SamLee-dedeboy/GreenMine}}, which implements a three-step prompting pipeline for the mining of DPSIR, and an interface with visualizations to support the interactive mining. The interface has two main views: \textit{Prompt View}, which supports interactive prompt refinement and execution, featuring an uncertainty chart for prompt evaluation; and \textit{DPSIR View}, which visualizes the mining results for collaborative discussions. \sam{
\marginnote{$\triangle$\_5\_1}All the LLM-related computation uses the ``gpt-4o-mini'' model from OpenAI, accessed through official API requests. 
}

\paragraph{Dataset and Segmentation}
The system supports datasets of interview transcripts.
Our targeting dataset consists of 19 transcripts of interviews that the collaborating experts have conducted with the residents of Lyudao. 
\sam{
\marginnote{$\triangle$\_5\_2}
To avoid reaching the context length limits, we first segment each transcript into ``snippets'', i.e., consecutive conversations about a specific topic.} 
To account for the variety of topics in the transcripts, the segmentation method is based on a prompt template, where the semi-structured interview questions are inserted, and the model is instructed to identify topics for segmentation. To avoid hallucinations (e.g., generating sentences not in the transcripts), the model outputs conversation indices, which are then used to segment the transcripts programmatically. Sanity checks are conducted to validate the result. This results in 598 snippets, with on average 6 conversations and 379 words (Chinese characters). 

\subsection{Prompting Pipeline}
Mining insights with the DPSIR framework is complex and requires expert validation, and it is unreliable to mine insights with a single prompt.
In response, we develop a prompting pipeline in three steps: \textit{Indicator Identification}, \textit{Variable Identification}, and \textit{Link Identification}, as shown in~\autoref{fig: pipeline}.
Conceptually, the prompts convert each mining task into a multi-label classification task, in which the model is instructed to choose all applicable labels from a candidate list (\textbf{R3}). 
Environmental experts can refine the candidate list (i.e., the indicators, variables, and their definitions) to fit the research goal and dataset context (\textbf{R1}). 

\paragraph{Indicator Identification} 
In this step, the prompt instructs the model to identify occurrences of indicators in each snippet. The prompt template takes the indicator definitions and snippet content as input and outputs a list of occurred indicators, as well as the evidence sentences and a textual explanation. Note that we use ``concept'' instead of ``indicator'' in the prompt, since ``indicator'' can mean differently in the model's training corpus.  

\paragraph{Variable Identification} Based on the assigned indicators, a second prompt template identifies the variables for each indicator in each snippet. Experts can edit the variable list, or refine the definition of each variable. 
The template takes three inputs:
(1) the indicator and the variable list (and definitions),
(2) the snippet content, and
(3) the textual explanation generated in the first step.
 The model outputs a list of occurred variables, as well as the evidence and explanation. Similarly, we use ``tag'' instead of ``variable'' in the prompt. 

\paragraph{Link Identification} Based on the identified variables, the relationships between all pairs of variables (if exist) are mined. We formulate it as a binary classification task on each pair of variables, where the model outputs (relationships) ``exist'' or `not exist''  given a pair of variables. This formulation increases the computation cost, but the results are more reliable due to shorter prompts and easier instructions. The template takes a pair of variables (and definitions) and the snippet content as input, and outputs the source, target, relationship, evidence, and explanation, or ``None'' if no relationship exists. To maintain the creativity of the model, we do not specify a list of predefined relationships in the prompt. During experiments, we observe some frequent relationships, such as ``positive/negative correlation'', ``causality'', ``interconnected'', or ``health-affecting'', showing that the model can identify and distinguish various kinds of relationships without human intervention.

\subsection{Uncertainty Estimation}
Unlike conventional text mining, there is often no labeled dataset available in prompt engineering. While some researchers have proposed to use pseudo-labels~\cite{malik2024pseudolabel}, it is still not feasible in DPSIR mining because the labels are incrementally enriched and contextualized as experts investigate the corpus. 

Given these unique characteristics, we draw inspiration from a recent work that estimates the uncertainty of LLM outputs by sampling multiple outputs for the same input~\cite{chen2024quantifyinguncertaitny}. 
\sam{
\marginnote{$\triangle$\_5\_3}Conceptually, we execute every prompt $k$ times and measure the consistency between outputs. 
Specifically for our prompting pipeline, since each step is formulated as a multi-label classification task (e.g., the response of \textit{Indicator Identification} can be any power set of $\{D, P, S, I, R\}$), we use set similarity metrics such as Jaccard Distance~\cite{jaccard1912jaccardsimilarity} to measure the consistency between $k$ response.
More generally, we recommend using information entropy~\cite{shannon1948informationentropy} for single-label classification tasks or semantic uncertainty~\cite{cheng2024relic} for unstructured outputs.
}

Using \textit{Indicator Identification} as an example, $A_{ij}$ denotes the identifed indicators for snippet $s_i$ in iteration $j \in (1, 2,...k)$, and $\forall a \in A_{ij}, a \in \{D, P, S, I, R\}$. Then we use average pairwise Jaccard Distance to compute the uncertainty $D_i$ for snippet $s_i$:
\vspace*{-0.1cm}
\begin{equation}
\begin{split}
    D_i = \displaystyle \sum_{1 \leq j_1 < j_2 \leq k}\frac{J(A_{ij_1}, A_{ij_2})}{k(k-1)/2}, \\
    where \ J(A, B) = \frac{|A \cup B| - |A\cap B|}{|A \cup B|}
\end{split}
\end{equation}
For variables, $a \in A_{ij}$ are the variables specified by the environmental experts in the form of strings. For links, we consider $a_1=a_2 \Leftrightarrow (src_1, dst_1)= (src_2, dst_2)$, where $(src, dst)$ are the variables, i.e., we do not consider the relationship when matching two links. 
Using this uncertainty estimation, we are able to support the evaluation of prompting performances on all subtasks (\textbf{R2}).

\sam{\marginnote{$\triangle$\_5\_4}Note that in \system, we set the GPT model's temperature to $0$ for maximal determinism. Still, uncertainty scores can reach $0.8$ in some cases, revealing significant response inconsistency and showing the necessity of consistency estimation. 
Although calculating uncertainty theoretically introduces up to $k$ times overhead by repeating each prompt $k$ times, we leverage OpenAI's extensive cloud resources to reduce the overhead, by sending all the API requests for one execution concurrently. Since OpenAI's computing clusters likely have the resources to handle all API requests simultaneously, the bottleneck of the system is simply Network I/O. 
In practice, uncertainty calculation adds only a $1.5x$ overhead when $k=5$.
}

\begin{figure*}[t]
     \centering
    \includegraphics[width=\textwidth]{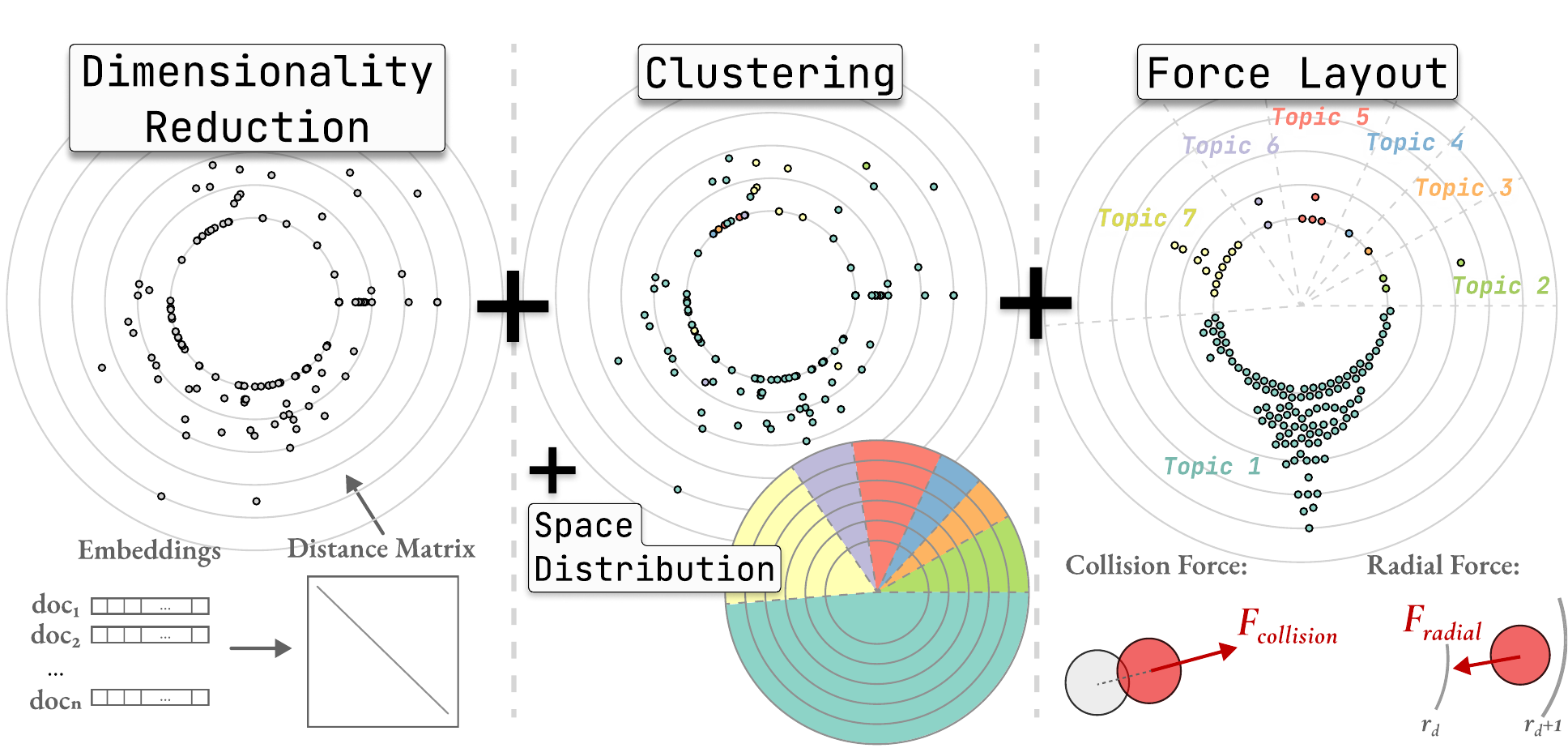}
    \caption{The process of generating the uncertainty chart. The dots are positioned using polar coordinates. The angles encode semantic similarity, and the radius encodes uncertainty. For angles, we start by generating embeddings from the snippets and calculate a semantic distance matrix using cosine distance. For different steps in the pipeline, we use the corresponding evidence sentences in each snippet to generate the embeddings. The distance matrix is optimized with the objective function defined in~\autoref{eq: dr}, resulting in angles for each snippet. The snippet embeddings are then clustered using agglomerative clustering and aggregated. The radial space is then divided proportionally by the cluster sizes. Then we employ a force-directed layout, with a collision force to avoid node overlapping, and a radial force that attracts nodes to their radius in the polar coordinate. The design considers the visual clarity of the layout while encoding uncertainty and semantics in one chart.}
    \label{fig: uncertainty_chart_improvements}
  \vspace*{-0.5cm}
\end{figure*}

\subsection{Uncertainty Chart}
\sam{
\marginnote{$\triangle$\_5\_5}Since the mining results can be overwhelming to make sense of and evaluate, we design an uncertainty chart (\autoref{fig: uncertainty_chart_improvements}) to organize the mining results with topics and uncertainties (\textbf{R2}).
}
 Each circle on the chart represents a snippet. The position $(\theta, r)$ is calculated in polar coordinates, where the angle $\theta$ encodes the semantics of the snippet's content and the radius $r$ encodes the uncertainty score. As a result, semantically similar snippets are placed together to form clusters, and the uncertain snippets are placed at the peripherals. 

Incorporating semantics into the visualization is a design choice motivated by the progressive characteristic of DPSIR text mining. 
In the beginning, users have a limited understanding of the corpus topics, and they would refine the DPSIR taxonomy as they explore. 
Once the taxonomy is refined, they would re-execute the prompt, evaluate its performance, and then learn more about the corpus.
The uncertainty chart design supports this interleaved sensemaking and evaluation (\textbf{R2}).
The chart supports two kinds of visual exploration. To find missing definitions for the current DPSIR taxonomy, users can investigate uncertain snippets, which are outliers in the peripheral region. To verify that the model's understanding is aligned with the expert's intent, users can investigate certain snippets, which are grouped in the center region.
Next, we introduce the semantic angle calculation and how the visual clarity of the chart is improved.

\subsubsection{Semantic Angle Calculation} 
\sam{
\marginnote{$\triangle$\_5\_6}
The semantic angle calculation combines document embedding models, dimensionality reduction, clustering, and force-directed layout (\autoref{fig: uncertainty_chart_improvements}). The goal is to find angles that maximally preserve the semantic similarities between snippets while reducing overlaps.
}
\paragraph{Capturing semantics}
We use document embeddings to capture the semantic meanings of each snippet. Instead of using the snippet conversations, we use the evidence sentences and explanations generated by LLMs as the embedded content. This way, the embeddings capture only the semantics relevant to the prompting instructions, e.g., sentences and explanations that suggest the occurrence of a \textit{Driver} when conducting indicator identification. We use OpenAI's ``text-embedding-3-small'' model with 1536 dimensions for this step.

\paragraph{Dimensionality reduction into radial space}
Core to the semantic angle calculation is preserving the high-dimensional distances in a single dimension. Different from a 1D t-SNE~\cite{van2008tsne} or MDS~\cite{cox2000mds}, the distances in the reduced dimension should be \textit{circular}, i.e., in the range of $[0, 2\pi)$, points at $\theta+\epsilon$ and $2\pi-\theta-\epsilon$ are of equal distance with points at $\theta$.
To achieve this, we formulate the reduction similar to MDS, but instead of mapping points into 2D or 3D space, we map them into a 1D circular space.
Given a distance matrix $D$, where $D_{ij}$ is the cosine distance between the embeddings of snippet $i$ and snippet $j$, the objective is formulated as:
\begin{equation}\label{eq: dr}
    \begin{split}
    \min_{\theta_1,\theta_2,\theta_3,...\theta_n}\displaystyle \sum_{i=1}^{n}\sum_{j=i+1}^{n}(d_{ij} - D_{ij})^2, \\
    d_{ij}=1 - cos(min(|\theta_i - \theta_j|, 2\pi - |\theta_i - \theta_j|)), \\
    \theta_i \in [0, 2\pi], \quad i = 1,2,3,\dots, n
    \end{split}
\end{equation}
We optimize for $\theta_i$, such that the squared difference between $d_{ij}$ and $D_{ij}$ is minimized. 
The calculation of $d_{ij}$ guarantees the circular characteristic and rescales to match $D_{ij}$. The objective is optimized with Limited-memory BFGS (L-BFGS)~\cite{liu1989limited}.
Next, we introduce how cluttering is reduced with clustering and force-directed layout.

\paragraph{Clustering}
We first apply Agglomerative Clustering~\cite{steinbach2000doccluster} on the embeddings using cosine similarity with a threshold of 0.5, which assigns a cluster to each snippet. With the clusters and angles from DR, we calculate the mean angle for each cluster and use the mean to sort the clusters, and then distribute the space in $[0, 2\pi]$ proportional to the cluster sizes. This maintains the proximity of semantically similar clusters. Since each cluster represents a topic, we want to support users to make sense of the topics at a glance. We use a prompt to assign topic labels to clusters using the snippet contents, which are attached to respective cluster regions in the chart.


\begin{figure*}[t]
    \centering
    \includegraphics[width=\textwidth]{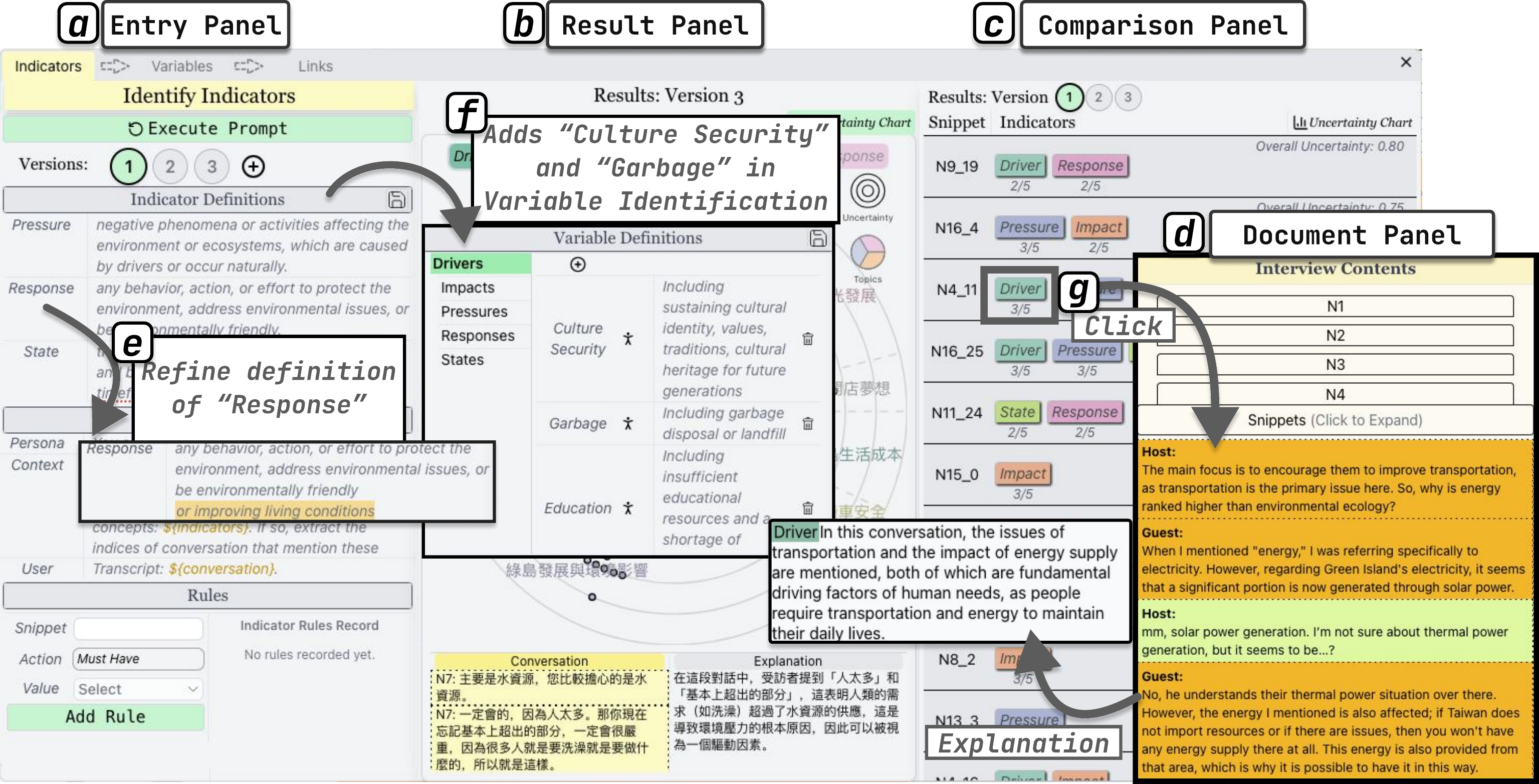}
    \caption{The prompting interface consists of three panels: (a) The Entry Panel, where users can adjust DPSIR taxonomy definitions and prompting templates. (b) The Result Panel, where users can choose to show the results in a list view or an uncertainty chart. (c) The Comparison Panel can be used to compare two prompts side-by-side. (d) The Document Panel shows the corpus contents. Using this interface, \textbf{E1} adds ``improving living conditions'' in the definition of response in (e), and adds ``culture security'' and ``garbage'' in the Driver variables in (f). (g) Users can also navigate to the evidence snippets, which are annotated with explanations generated by LLMs. }
    \label{fig: case_study}
    \vspace*{-0.6cm}
\end{figure*}

\paragraph{Force-Directed Layout}
Finally, we use a force-directed layout~\cite{fruchterman1991forcelayout} to reduce node overlapping.
At rendering time, all circles are initialized with their coordinates $(\theta_i, r_i)$. 
We apply two forces on the nodes: (1) a collision detection force that repels overlapping nodes, (2) a radial position force that attracts each node to its radius. We also implement clipping to limit the nodes in their allocated cluster spaces. 
\sam{\marginnote{$\triangle$\_5\_7}Although force-directed layout might trade accuracy for visual clarity, it is worth noting that the chart construction process provides relatively large flexibility in node placement: First, the uncertainty score (main encoding of the chart) is encoded by the radius of the polar coordinate, so nodes can be placed at varying angles without changing the uncertainty score encoding. By incorporating the radial position force, the layout prioritizes reducing overlaps with angles rather than radius. Second, semantics of the snippets (encoded by angles) is inherently vague, and it is less likely to cause confusion when the angles are adjusted.} Together, node overlapping is reduced with minimum adjustments to coordinates. 

\vspace*{-0.1cm}
\subsection{Prompting Interface}
The prompting interface consists of three panels: an \textit{Entry Panel} for users to edit DPSIR taxonomy and prompts; a \textit{Result Panel} where users can inspect the prompting results using the uncertainty chart or a list view; and a \textit{Comparison Panel} where users can select a previous version of prompting result side-by-side.
\sam{
\marginnote{$\triangle$\_5\_8}All three steps in the pipeline have an Entry Panel, a Result Panel, and a Comparison Panel. On top of that, some UI adjustments are made to better support each step, e.g., the list view for \textit{Link Identification} provides the Link Graph. 
}
Next, we introduce each panel and its interactions.

\begin{figure*}[t]
    \centering
    \includegraphics[width=\textwidth]{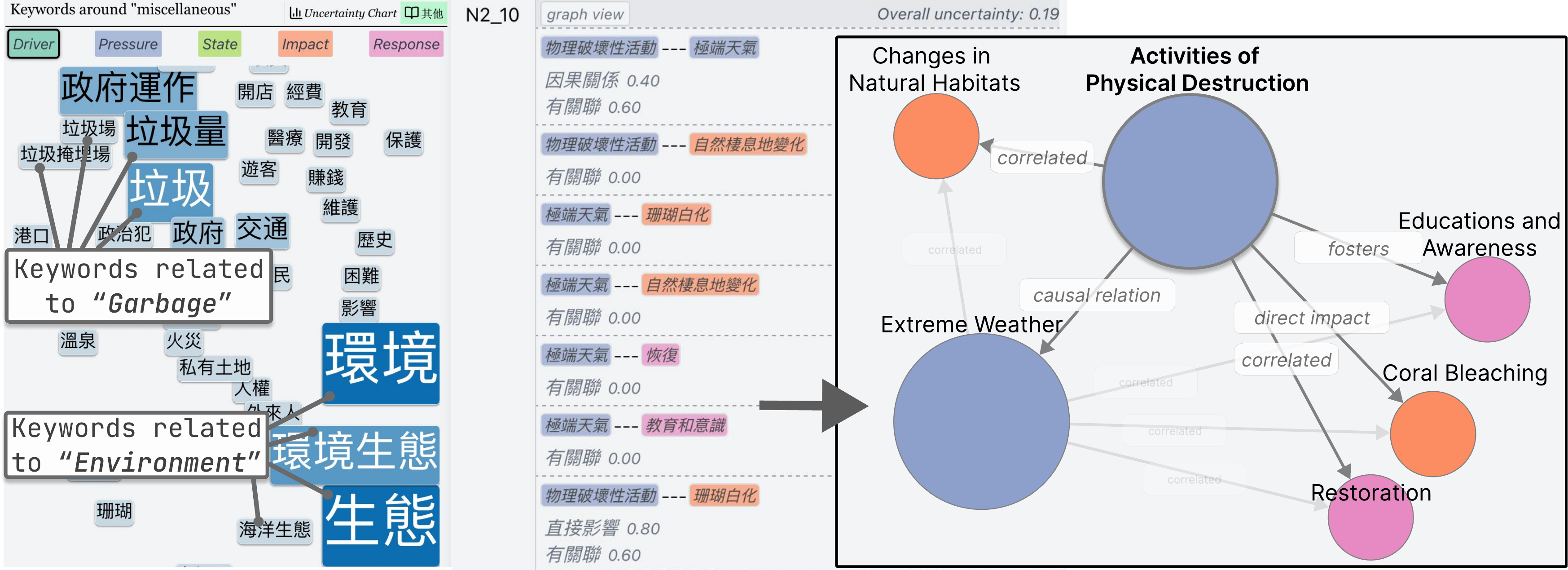}
    \caption{\textbf{Left}: Keyword cloud designed to support the exploration of ``miscellaneous'' variables. The keyword cloud combines embeddings and dimensionality reduction to position semantically similar keywords closer. It shows that ``garbage'' and ``environment'' are two significant keywords in the miscellaneous variables for Drivers. \textbf{Right}: The link graph design helps users make sense of complex links in a snippet. The graph shows that ``Activities of Physical Destructions'' is correlated with ``Changes in Natural Habitats'', ``Coral Bleaching'', and other variables.}
    \label{fig: simple_vis}
    \vspace*{-0.5cm}
\end{figure*}

\vspace*{-0.15cm}
\subsubsection{Entry Panel}
The Entry Panel supports four functionalities: version control, DPSIR taxonomy input, prompt input, and rule input (\autoref{fig: case_study}-a). The panel supports managing different versions of inputs for each step. Each version has a set of DPSIR definitions, prompts, and the last execution result. 
The versions are also necessary when specifying data dependency between steps in the pipeline.
\vspace*{-0.1cm}
\paragraph{Taxonomy and Prompt Input}
In the first two prompting steps, \textit{Indicator Identification} and \textit{Variable Identification}, users can edit the indicator or variable definitions in natural language, add or remove variables, or specify the variable type (societal or environmental). 
Each prompting step has a unique template, in which the dataset and application context can be specified. Each template is divided into three sections: \textit{Persona}, \textit{Context}, and \textit{User}. This section division follows the state-of-the-art prompting methodologies~\cite{promptengineeringuide} and helps users make sense of the template. 
Users can use a special ``\$\{\}'' tag to specify a variable insertion into the template. 
These specifications are inserted into the prompt templates automatically.

\vspace*{-0.1cm}
\paragraph{Rule Entry} The rules enable users to manually modify prompting results and automatically apply them to all versions of results. Each rule should specify a target snippet, a condition (``must'' or ``must not'' have), and a value, which can be indicators, variables, or links. Note that the rules are not used in the prompts or as few-shot examples. We implement this feature for users to cover corner cases where the model fails to capture some nuances.

\vspace*{-0.15cm}
\subsubsection{Result Panel and Comparison Panel}
Result and Comparison Panel (\autoref{fig: case_study}-b, c) support the evaluation of prompt performances (\textbf{R2}) by showing the results using the uncertainty chart or a list view.
The Result Panel always displays the current version, while the Comparison Panel allows selecting any version. We also designed two visualizations—Keyword Cloud and Link Graph—to aid sensemaking in variable and link identification.

\vspace*{-0.1cm}
\paragraph{List View}
The list view shows all results sorted by uncertainty (high to low). By clicking on a result, user can quickly navigate to the evidence and explanations in the Document Panel (\autoref{fig: case_study}-g). 

\vspace*{-0.1cm}
\paragraph{Keyword Cloud}
During variable identification, we reserve a label for ``miscellaneous'' to cover everything that the experts have not considered. Adding new variables by iteratively investigating the ``miscellaneous'' variables is the core of refining the DPSIR variable lists. To support this, we aggregate the keywords from the evidence sentences for results containing ``miscellaneous'', and implement a semantic word cloud~\cite{xu2016semanticwordcloud} using embeddings and Kernel PCA~\cite{scholkopf1997kernel}, as shown in~\autoref{fig: simple_vis} (left).
\sam{
\marginnote{$\triangle$\_5\_9}Users can switch to show the keyword clouds by clicking the button located at the top right of the result panel for \textit{Variable Identification}.
}
We use color and size to double-encode the word frequencies, and use proximity to encode semantic with cosine similarity.
The design also considers maintaining the visual continuity between multiple rounds of iteration.
Specifically, we want the semantic landscape to remain consistent.
Thus, we need a \textit{parametric} dimensionality reduction method, where an explicit mapping function (i.e., projection to low-dimensional space) can be reused, excluding popular non-parametric methods like t-SNE~\cite{van2008tsne} or MDS~\cite{cox2000mds}.
Kernel PCA not only satisfies this constraint, but also captures the non-linear semantic relationships. Similar to the uncertainty chart, we use a collision force with rectangular overlap detection to prevent keyword overlapping.

\vspace*{-0.1cm}
\paragraph{Link Graph}
During link identification, the system supports the visualization of the result for a snippet in a node-link diagram (\autoref{fig: simple_vis}-right), which makes it easier to understand the result and identify graph-related patterns than in a list. Each node is a variable colored by their indicator types. Node radius encodes degree (frequency), and link opacity encodes uncertainty. The links are annotated with the relationship label extracted by the model, and users can click on the links to navigate to the evidence in the Document Panel. Since the graph size is typically small, we use a force-directed layout and implement node dragging in case the labels are occluded.

\subsection{Visualizing links in DPSIR Graph}
\begin{figure*}[t]
    \centering
    \includegraphics[width=\textwidth]{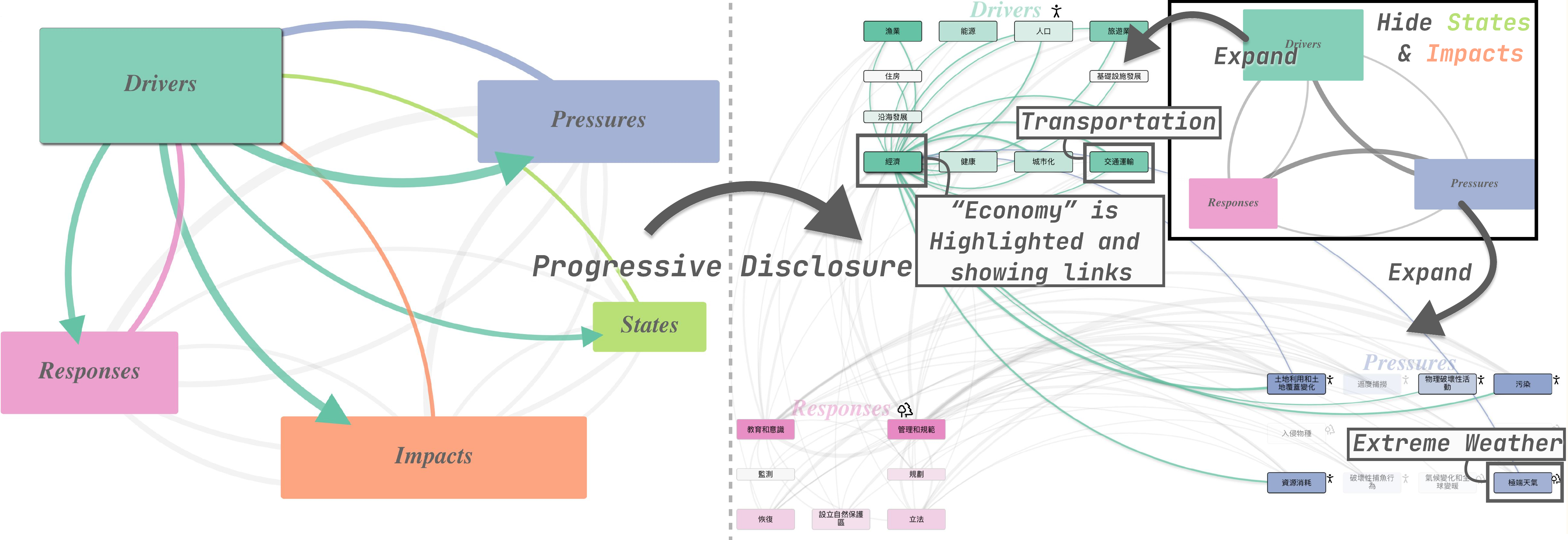}
    \caption{\textbf{Left}: The DPSIR Graph shows the aggregated mining results in a progressive graph. The design follows the typical DPSIR diagram in environmental studies. The Driver block is highlighted with ingoing and outgoing links colored by the source of the link. \textbf{Right}: The States and Impacts are hidden, and the rest of the indicators are revealed to show the variables and their links. \textbf{E1} found that contrary to the literature, ``Economy'' and ``Transportation'' are two significant drivers with many linkages, and ``Extreme Weather'' is a significant pressure. }
    \label{fig: DPSIR}
    \vspace*{-0.5cm}
\end{figure*}

During the mining tasks, the system prioritizes to visualize only a selective subset of results. After the mining, the system can visualize all mined links in the DPSIR Graph, where each node is a variable. 
As shown in~\autoref{fig: DPSIR}, the DPSIR Graph is designed specifically for the DPSIR framework, extending common visualizations created in the environmental science community~\cite{atkins2011dpsir}, featuring a radial layout and progressive disclosure support. It is designed for quick validations and sensemaking during a collaborative discussion (\textbf{R4}). 

\vspace*{-0.1cm}
\paragraph{Responsive Radial Layout} 
Using the cyclic characteristic of the DPSIR framework, the graph is organized in a radial layout. The design follows existing diagrams in environmental studies~\cite{atkins2011dpsir} and emphasizes simplicity and interactivity to better support collaborative discussion.
Each indicator is assigned a unique color consistent with the rest of the system. To lay out each block of indicator, we divide $[0, 2\pi]$ proportionally to the total degree of variables in each indicator (D, P, S, I, R), and put the block at the center of the allocated space. In this formulation, the system can support users to exclude irrelevant indicators during discussion sessions. 
For example, users can hide \textit{States} and \textit{Imapcts} if the discussion is not focusing on these indicators (\autoref{fig: DPSIR}-right), and the layout will automatically redistribute the space to place each block evenly.

\vspace*{-0.1cm}
\paragraph{Progressie Disclosure} 
 Each block of indicators can be interactively ``opened'' to reveal the variables (\autoref{fig: DPSIR}-right), organized in a squared layout to save the central space for internal links. The color saturation of each variable encodes degree (frequency of occurrence). To prevent external links from concentrating in a small space, we put the four highest-degree variables at the four corners of each block. This also helps users locate the most significant variables. The link width and opacity between each variable double-encode the intensity (frequency of the link). Link color is determined by the source indicator type.
  Users can click on any variables or links to highlight them, and navigate to the evidence in the Document Panel.
 The interaction design allows users to focus on relevant subsets of the DPSIR graph during discussions, 
 and seamlessly navigate to supporting evidence in the documents for detailed understanding.

\vspace*{-0.15cm}
\section{Case Study}
In this section, we introduce a case study in which \textbf{E1}, a researcher in environmental science, uses the system to refine the DPSIR taxonomy. \textbf{E1} has collected an initial set of DPSIR taxonomy from a literature review. He has imported this initial taxonomy into the system and done the first extraction (\autoref{fig: case_study}-a). Next, he inspects the results using the uncertainty chart to refine the taxonomy. 

\begin{figure}[]
     \centering
    \includegraphics[width=\columnwidth]{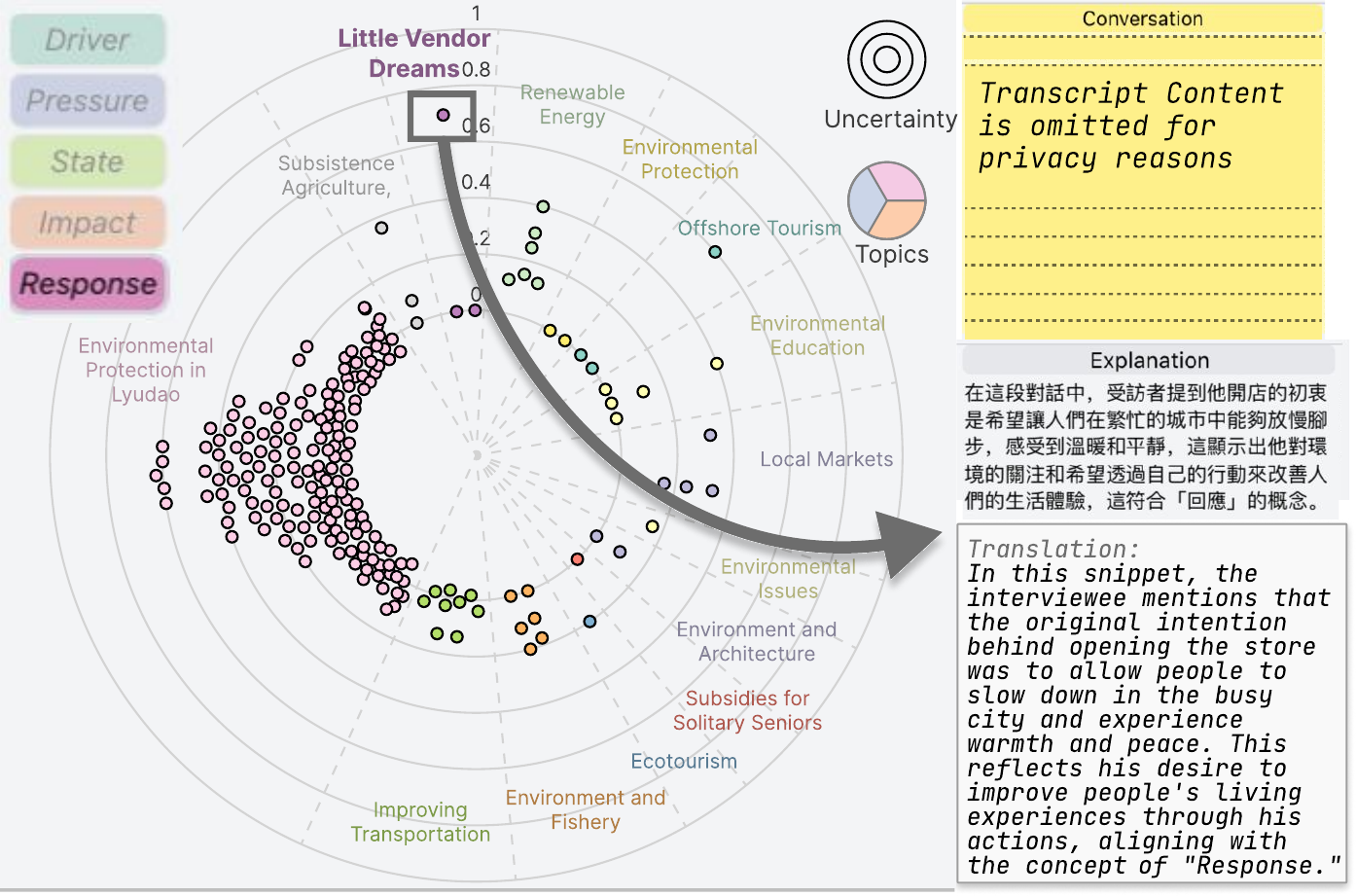}
    \caption{An uncertainty chart generated in the case study.
    Each dot in the chart represents a snippet that mentions the indicator ``Response'', positioned with polar coordinates. 
    From the chart, most topics fit \textbf{E1's} expectation except a snippet under ``Little Vendor Dream'' with high uncertainty. He clicks the snippet to inspect the relevant conversations and the LLM-generated explanation. 
    He finds that the snippet has high uncertainty because the Response definition is incomplete, 
    so he refines the definition accordingly. 
    }
    \label{fig: uncertainty_chart}
  \vspace*{-0.6cm}
\end{figure}

\vspace*{-0.12cm}
\paragraph{\textbf{\textit{Refining definition of \textit{Response}}}} The initial definition of Response is ``any behavior, action, or effort to protect the environment, address environmental issues, or be environmentally friendly''. 
\sam{\marginnote{$\triangle$\_6\_1}The uncertainty chart (\autoref{fig: uncertainty_chart}) organizes the results mentioning ``Response'' by topics and uncertainties, making it easier to make sense of and identify outliers.
}
\textbf{E1} sees that most topics fit this description and have low uncertainty. An exception is ``Little Vendor Dream'', which has a node with $0.7$ uncertainty, and the topic does not suggest something that is typically considered as ``Response''. Upon closer inspection, the snippet describes the interviewee's dream of establishing an eco-friendly, and serene shop, where visitors can escape the hustle and bustle of big cities and experience the warmth and tranquility that only nature can offer. This snippet has high uncertainty because it only roughly fits the current definition of Response. \textbf{E1} believes this should be included, and adds ``improving living conditions'' to the definition of Response.

\vspace*{-0.12cm}
\paragraph{\textbf{\textit{Enriching variables in \textit{Driver}}}}
During the refinement of Driver variables (\autoref{fig: case_study}-f), \textbf{E1} finds a snippet with high uncertainty about the White Error, which is a period of political repression from 1947 to 1987 that disrupted many tribal cultures. As a result, Lyudao has a district dedicated to commemorating this event. 
\textbf{E1} believes this is a driver that is often overlooked by the environmental science literature but is crucial to consider when making policies for Lyudao. 
\textbf{E1} also observes from the keyword cloud for ``miscellaneous'' that ``garbage'' frequently occurs (\autoref{fig: simple_vis}, left). Upon closer inspection, he finds that garbage disposal is a serious issue for the residents of Lyudao.
Based on these two observations, \textbf{E1} adds ``culture security'' and ``garbage'' to the variables of drivers. 

\vspace*{-0.12cm}
\paragraph{\textbf{\textit{Investigating links}}}
While refining the DPSIR taxonomy, \textbf{E1} makes several insights from the transcripts. For example, he finds a snippet about the coral reefs on Lyudao (\autoref{fig: simple_vis}, right). The link graph shows that there are activities of physical destruction that cause changes in natural habitats. He clicks the link to inspect the conversation and finds an interviewee's response that mentions the usage of boat anchors causing physical damage to the underwater corals. The links also reveal that the interviewee calls for restoration activities and education to raise environmental awareness. 

\vspace*{-0.12cm}
\paragraph{\textbf{\textit{Insights from the DPSIR Graph}}}
After various refinements to the indicator and variable definitions, \textbf{E1} observes some interesting patterns from the DPSIR graph (\autoref{fig: DPSIR}-right). For example, contrary to the existing literature, ``economy'' and ``transportation'' are two significant \textit{Drivers} with many occurrences in the context of Lyudao. Similarly in \textit{Pressure}, the interviewees express greater concern about ``extreme weather'' than ``climate change''. \textbf{E1} hypothesizes that this is due to the more immediate and severe impact on their livelihoods. This observation suggests that the infrastructure and economy in Lyudao are underdeveloped and vulnerable to ecosystem changes, and the environmental policies must consider that. Other insights confirm his hypotheses, such as the objection to thermal power plants or the need for better tourism management.

\vspace*{-0.15cm}
\section{Expert Review}
To better understand the potential and limitations of our approach, we demonstrated the system to experts participating in the Lyudao project.
In addition to \textbf{E1} and \textbf{E2}, we invited three more experts and host a discussion session. \textbf{E3} is a senior researcher in environmental science and is in charge of the interview that \textbf{E1} analyzed. \textbf{E4} is a research assistant with a Master's degree in environmental science, and \textbf{E5} is a PhD student in earth science. None of the three experts had seen the system before the discussion.

The procedure started with a brief introduction of the overarching system goal and the interface design. Then, \textbf{E1} presented his findings from the usage of the system, using both the uncertainty charts and the DPSIR Graph. During the presentation, we encouraged all experts to ask questions and provide their thoughts. The discussion lasted over 2 hours and was video-recorded. Below, we summarize the expert feedback and discuss the implications.

\vspace*{-0.1cm}
\paragraph{System design is well-received}
The system design is clear and intuitive. Despite lacking technical knowledge in prompt engineering or text mining, the experts found the three-step mining pipeline sensible and easy to evaluate. They easily understood the uncertainty chart's visual encodings. During \textbf{E1's} presentation, he effectively used the DPSIR Graph without external tools, demonstrating its suitability for collaborative discussions. Additionally, \textbf{E1}'s week-long usage without issues highlights the usability of the prompting functionality and uncertainty chart.

\vspace*{-0.15cm}
\paragraph{Complementing literature review}
All experts agreed that the system is an effective tool to provide unexpected insights after seeing \textbf{E1's} findings. \textbf{E4} was especially fond of the automatic support, knowing the time and effort needed to manually analyze these transcripts. \textbf{E3} agreed that \textit{``Looks like we are scattering the transcripts and then reorganizing them in response to our research questions with better and clearer definitions (compared to literature review).''}
\textbf{E1} summarized from his experience that computational extraction reduces the cognitive bias that human experts might have from literature knowledge.
This shows the system not only streamlines transcript analysis but also mitigates cognitive biases, improving the objectivity and precision of the derived insights.

\vspace*{-0.15cm}
\paragraph{Informing policymaking}
During the discussion, \textbf{E2} explored how the DPSIR Graph could guide policymaking, focusing on \textit{Response} variables to gauge public opinion on current policies. \textbf{E1} agreed, suggesting the inclusion of other relevant variables: \textit{``For example, for fishery policies, we can identify connected variables, indicating where policies should focus ({E1}).''} This highlights the DPSIR Graph’s potential to inform policy by identifying key variables and incorporating public opinion for data-driven decisions.

\vspace*{-0.15cm}
\paragraph{Facilitating public-facing presentation}
\textbf{E3}, leading interview recruitment, was keen on improving public outreach with the system, noting that Lyudao residents historically distrust government-funded research. \textbf{E2} agreed, stating, \textit{``Residents want to feel heard and see their opinions considered. This tool could help demonstrate that.''} Still, \textbf{E3} emphasized the need to simplify the DPSIR Graph to match the public's visualization literacy. Balancing these views, the system shows potential as a bridge between research and the public.

\vspace*{-0.15cm}
\paragraph{Comparing detailed research questions}
While the system is designed for exploratory purposes in mind, it could also be applied to study and compare concrete research questions. As suggested by \textbf{E3}, users can give very narrow indicator definitions and short lists of variables to focus on only one research question: \textit{``We can first set a few research questions, for example, pressures on oil and gas, or pressures on the fishery, and then adjust the definitions accordingly. We can even compare different versions of definitions, which might reveal more focused insights. Such insights would also be more suitable for the public to digest.''}
This adaptability highlights the system's potential to support detailed analyses.
\section{Limitations}
While \system \ has proven effective for our collaborating experts, we discuss limitations that could constrain its applicability.
\sam{
\paragraph{Incorporating multiple sources of data}
\marginnote{$\triangle$\_8\_1}The system currently supports only a specific data format for interview transcripts, which is quite limited given LLMs' ability to handle various formats. Extending support for other formats require updates to several system modules, such as the segmentation module and Document Panel.
}
\paragraph{Computation Scalability}
The system is limited in computational scalability. 
The dataset on Lyudao is segmented into 598 snippets with an average length of 379 Chinese characters, which is not a large dataset. During the mining, we observed an average runtime of 150 seconds for indicator and variable identification respectively, 
\sam{\marginnote{$\triangle$\_8\_2}}and 600 seconds for link identification. \sam{Note that in our implementation, the LLM is accessed through network APIs and optimized with multithreading to reduce uncertainty calculation overheads. For local LLM inferences, the overheads could not be reduced and the runtime would drastically increase. 
}
\paragraph{Visual scalability}
The uncertainty chart and DPSIR graph face visual scalability issues. \sam{
\marginnote{$\triangle$\_8\_3}
With nodes clustering in narrow ranges or with excessive topics, the force directed layout might push nodes too far away from their original positions, affecting the accuracy of the uncertainty encoding. 
The DPSIR graph relies on interactions for clarity as the number of nodes increases. 
While suitable for collaborative discussions, it may not work well for static displays like posters, limiting its applicability to larger datasets or more display environments.
}
\sam{
\paragraph{Limitations in evaluation}
\marginnote{$\triangle$\_8\_4}The system's evaluation is based solely on feedback from a small group of collaborating experts, which may not be comprehensive. While feedback has been positive, the experts received a detailed tutorial from the authors. Without the tutorial, users unfamiliar with LLMs or advanced visualization—likely common in the environmental science community—may find the system challenging to use. Additionally, the system has not been tested with other LLMs, making it unclear if model choices would impact the outcome and overall performance.

\paragraph{Ethical considerations}
\marginnote{$\triangle$\_8\_5}Finally, potential ethical concerns~\cite{weidinger2022llmrisk}, such as privacy and biases, remain unaddressed. During \system's development, our collaborating experts raised data privacy concerns, prompting us to de-identify interview transcripts. While this is standard for transcripts, similar standards may not exist for other formats like field reports, leaving privacy a concern. Additionally, LLMs may exhibit biases from their training corpus, potentially undermining their ability to fairly extract insights from large datasets.
}
\sam{
\section{Implications beyond Environmental Study}
}
During the development of \system, we addressed several technical challenges that may also exist in other application scenarios. In this section, we discuss the lessons learned that inform visual analytics system developers beyond environmental study. 

\paragraph{Uncertainty evaluation for knowledge-intensive tasks}
Our human-in-the-loop evaluation using uncertainty has proven effective for DPSIR mining tasks. While LLMs perform well on general tasks, their capabilities on knowledge-intensive tasks remain limited due to insufficient training data and the complexity of eliciting knowledge as prompts~\cite{harvel2024llmknowledge}. By leveraging uncertainty estimation, knowledgeable users can iteratively refine prompts until uncertainty is minimized. 
Integrating evaluation with exploration aids this process, providing actionable insights for prompt refinement. Thus, the uncertainty chart also suits other knowledge-intensive applications.

\paragraph{Progressive taxonomy construction in thematic analysis}
The data mining in the DPSIR framework resembles that of thematic analysis, facing similar technical challenges. In thematic analysis, analysts begin with an initial codebook and progressively refine it~\cite{fereday2006thematic}. Some researchers have used LLMs to aid this process~\cite{dai2023llmintheloop, yan2024chatgptthematic}, but they encounter reliability and consistency issues due to inadequate feedback loops. Our approach addresses this by combining uncertainty evaluation and topic exploration in the feedback loop, offering validation and actionable feedback. The insights from our work can improve LLM-facilitated thematic analysis, particularly in enhancing reliability and consistency.

\sam{
\paragraph{Incorporating semantic uncertainty}
For broader applicability, we can use semantic uncertainty to incorporate free-form mining results.
\marginnote{$\triangle$\_9\_1}
In our work, we have demonstrated the benefit of using uncertainty metrics to support users evaluate LLM responses.
To adapt to scenarios beyond classifications, e.g., in extracting facts from a corpus, 
}the uncertainty chart can be extended with semantic uncertainty~\cite{kuhn2023semanticuncertainty, cheng2024relic}, which uses linguistic variances to measure the uncertainty of textual responses. 
With this, the uncertainty chart can be extended to more mining scenarios, providing robust uncertainty evaluation support for both structured and unstructured responses.

\section{Conclusion}
In this paper, we present \system, a system designed to facilitate progressive taxonomy construction based on the DPSIR framework for environmental studies. The system supports the interactive execution of a three-step prompting pipeline, where domain specifications can be inserted in prompts by experts. To support the evaluation of the prompts, we introduce an uncertainty chart that visualizes corpus topics and prompt output consistency. The uncertainty chart, along with other visualizations, supports interleaved evaluation and exploration to provide actionable feedback. Our proposed solution and the lessons learned offer valuable insight into supporting human-in-the-loop text mining beyond environmental studies.

\newpage

\acknowledgments{
This research is supported in part by the National Science Foundation via grant No. IIS-2427770, and by the University of California (UC) Multicampus Research Programs and Initiatives (MRPI) grant and the UC Climate Action Initiative grant.
}

\bibliographystyle{abbrv-doi}
\balance
\bibliography{template}
\end{document}